\documentclass[a4paper,11pt]{article}
\pdfoutput=1 

\usepackage{jheppub} 

\usepackage[T1]{fontenc} 

\title{\boldmath Confinement-Higgs Phase Crossover as a Lattice Artifact in $1+1$ Dimensions}

\author[]{Axel Cort\'es Cubero }

\affiliation[]{ SISSA and INFN, Sezione di Trieste; via Bonomea 265, 34136 Trieste, Italy.}

\emailAdd{acortes@sissa.it}

\abstract{We examine the phase structure of massive Yang-Mills theory in 1+1 dimensions. This theory is equivalent to a gauged principal chiral sigma model. It has been previously shown that the gauged theory  has only a confined phase, and no Higgs phase in the continuum, and at infinite volume. There are no massive gluons, but only hadron-like bound states of sigma-model particles. The reason is that the gluon mass diverges, being proportional to the two-point correlation function of the renormalized field of the sigma model at $x=0$. We use exact large-$N$ results to show that after introducing a lattice regularization and typical values of the coupling constants used in Monte Carlo simulations, the gluon mass becomes finite, and even sometimes small. A smooth crossover into a Higgs phase can then appear. For small volumes and large $N$, we find an analytic expression for the gluon mass, which depends on the coupling constants and the volume. We argue that this Higgs phase is qualitatively similar to the one observed in lattice computations at $N=2$.
}

\newcommand{\beq}{\begin{eqnarray}}
\newcommand{\eeq}{\end{eqnarray}}

\begin{document} 
\maketitle
\flushbottom

\section{Introduction}

Yang-Mills theories coupled to a matrix-valued Higgs field are known to have a confined phase and a Higgs phase in 2+1 and 3+1 dimensions \cite{fradkin}. The phase diagram in 1+1 dimensions is much less understood.

The author and P. Orland proposed studying the (1+1)-dimensional theory as a gauged principal chiral sigma model (PCSM) \cite{massiveyangmills}. We found that in the continuum theory at infinite volume, there is only a confined phase. The physical excitations are hadron-like bound states. We computed analytically the spectrum of the meson-like hadrons in the nonrelativistic limit, using knowledge of the exact S-matrix of the PCSM.

A numerical lattice study of the phase diagram of the same theory, with gauge group ${\rm SU}(2)$, was published by S. Gongyo and D. Zwanziger \cite{gongyozwanziger}. In this reference, they found evidence for both the confined and the Higgs phase, even in 1+1 dimensions. There seems to be a crossover between the two phases, instead of a sharp phase transition. This appears at first sight contradictory to the results of \cite{massiveyangmills}; however, what they found is that the crossover seems to disappear as the  volume of the system is increased. Their calculations suggest that there is a Higgs phase, but it disappears as they move towards infinite volume. In fact, within the paper, the authors make the statement, ``from this data at finite volume we cannot conclude that there is a phase transition to a symmetry-breaking [Higgs] phase at infinite volume". Furthermore, as we will suggest, the value of gluon mass monotonically grows as the continuum limit is taken.

Inspired by the nontrivial results of \cite{gongyozwanziger}, we extend the approach started in \cite{massiveyangmills} to study 
 analytically the (1+1)-dimensional theory with a lattice discretization and at finite volume. Our analytic computation is so far only possible at large-$N$, so it cannot be directly compared to the $N=2$ results of \cite{gongyozwanziger}. Nevertheless, we will see that the behavior of the Higgs phase we find at large $N$ is qualitatively similar to the $N=2$ results of \cite{gongyozwanziger}.
 
The mathematical object we study is the two-point correlation function of the renormalized-field operator of the PCSM. This function has been calculated at infinite volume in \cite{renormalizedfield}, \cite{asymptoticfreedom}, and finite volume in \cite{thooftthermodynamics}, using the LeClair-Mussardo formula for integrable field theories \cite{leclairmussardo}. 

Our main claim is that even though there is only a confined phase in the continuum theory at infinite volume, a Higgs phase can arise once a lattice regularization is introduced, particularly at very small volumes. 
Our calculation is valid only for large $N$, since the nonperturbative results needed are only available in this limit. We will later argue that a similar phenomenon can occur at finite $N$ (particularly at $N=2$, which is desired for comparison with numerical results), however, we can only rely on perturbative calculations, and only study the infinite-volume limit.

The PCSM has the action
\beq
S_{\rm PCSM}=\int d^2 x\,\frac{1}{2g_0^2}{\rm Tr}\partial_\mu U^\dag(x)\partial^\mu U(x),\label{pcsmaction}
\eeq
where $U(x)\in {\rm SU}(N)$. This model has been shown to be integrable, and its exact S-matrix is known \cite{wiegmann}. The action (\ref{pcsmaction}) has an ${\rm SU}(N)\times {\rm SU}(N)$ global symmetry given by $U(x)\to V_L U(x) V_{R}$, with $V_{L,R}\in {\rm SU}(N)$. The PCSM is asymptotically free and has a mass gap, which we call $m$ \cite{unsal}. 

One can obtain the massive Yang-Mills action by promoting one of the ${\rm SU}(N)$ symmetries of (\ref{pcsmaction}) to a local gauge symmetry. We choose to gauge the left handed symmetry, $V_L\to V_L(x)$. We introduce the covariant derivative $D_\mu=\partial_\mu-i\,eA_\mu(x)$, and field strength $F_{\mu\nu}=\partial_\mu A_\nu-\partial_\nu A_\mu-ie[A_\mu,A_\nu]$. The gauge field, $A_\mu(x)$ transforms under the gauge symmetry as $A_\mu(x)\to V_L^\dag(x) A_\mu(x) V_L(x)-\frac{i}{e} V_L^\dag(x) \partial_\mu V_L(x)$. The gauged PCSM action is 
\beq
S=\int d^2x\left[-\frac{1}{4}{\rm Tr}F_{\mu\nu}F^{\mu\nu}+\frac{1}{2g_0^2}{\rm Tr}(D_\mu U)^\dag D^\mu U\right].\label{gaugedpcsm}
\eeq

In Reference \cite{massiveyangmills}, it was argued that the model with action (\ref{gaugedpcsm}) is always in a confined phase, rather than a Higgs phase. The argument was made by examining this action in the axial gauge $A_1=0$. In this gauge, the action (\ref{gaugedpcsm}) becomes
\beq
S=\int d^2x\left[\frac{1}{2}{\rm Tr}(\partial_1 A_0)^2+\frac{1}{2g_0^2}{\rm Tr}(\partial_0U^\dag+ieU^\dag A_0)(\partial_0U-ieA_0 U)-\frac{1}{2g_0^2}{\rm Tr}\partial_1U^\dag \partial_1 U\right].\nonumber
\eeq
The gauge field $A_0$ can now be integrated out, obtaining the action
\beq
S=\int d^2x\left(\frac{1}{2g_0^2}{\rm Tr}\partial_\mu U^\dag \partial^\mu U+\frac{1}{2}j_{0\,a}^L\frac{1}{-\partial_1^2+e^2/g_0^2\,U^\dag U}j_{0\,a}^L\right),\label{integratedgauge}
\eeq
where $j_{\mu}^L(x)_b=-i{\rm Tr} t_b\partial_\mu U(x) U^\dag(x)$ is the Noether current associated with the left handed symmetry, and $t_b$, with $b=1,\dots,N^2-1$, are the generators of ${\rm SU}(N)$. The action (\ref{integratedgauge}) describes a PCSM with an additional potential between the left-handed charges.

Looking at (\ref{integratedgauge}) one would naively conclude that, since $U^\dag U=1$ (because $U$ is unitary), the potential between the charges is screened by a force carrying excitation of mass $e/g_0$. This would be interpreted as the massive gluon of the Higgs phase. As was discussed in Ref. \cite{massiveyangmills}, this reasoning is wrong. The main problem is that even though the bare field $U(x)$ is unitary, the physical renormalized field, which we call $\Phi(x)$, is not \cite{renormalizedfield}. The relation between the two fields is given by 
\beq
\langle0\vert{\rm Tr}\Phi(x)\Phi(0)^\dag\vert0\rangle=Z[g_0(\Lambda),\Lambda]^{-1}\langle0\vert{\rm Tr} U(x)U(0)^\dag\vert0\rangle,\label{definez}
\eeq
Where $Z[g_0(\Lambda),\Lambda]$ is a renormalization constant, and $\Lambda$ is the momentum cutoff of the theory. Because of the asymptotic freedom of the PCSM, the renormalization constant vanishes logarithmically with the cutoff. This means that renormalization pushes the mass of the gluon to infinity. Therefore one does not see any massive gluons in the physical spectrum of the theory. The model is actually in a confined phase.

Using the fact that for the bare field, $U^\dag(0)U(0)=1$, the renormalized gluon mass, $\mathcal{M}$,  is given by
\beq
\mathcal{M}=\frac{e}{g_0}Z[g_0(\Lambda),\Lambda]^{-1/2}=\frac{e}{g_0}\left[\frac{1}{N}\langle 0\vert{\rm Tr}\Phi(0)\Phi(0)^\dag\vert0\rangle\right]^{1/2}.\label{gluonmass}
\eeq
The only reason that there is no Higgs phase is that the PCSM field-renormalization constant vanishes as we increase the cutoff.  Consequently, the gluon mass diverges. It is then clear that once a cut-off is introduced by placing the theory on a lattice, the gluon mass becomes finite. In this letter we propose that placing the model in a small  volume can further suppress the divergence. Given the values of the couplings, $e,\,g_0$ used in \cite{gongyozwanziger}, the mass becomes small enough to be measured in Montecarlo simulations. We show that the dependence of the gluon mass on the volume agrees qualitatively with the results of \cite{gongyozwanziger}.

\section{Infinite volume PCSM correlation function} 

An expression for the two-point function of the renomalized field of the PCSM was found in Ref. \cite{renormalizedfield}. This expression was found using the fact that the PCSM is integrable, combined with the large-$N$ limit. The approach is to first find the exact S-matrix, and all the form factors (matrix elements of local operators) of the renormalized field using the integrable bootstrap program \cite{bootstrap}. Once all these form factors are known (they are known for the PCSM only in the large-$N$ limit, with $g_0^2N$ fixed), the two-point function is given by the spectral sum,
\beq
\mathcal{W}(x)&=&\frac{1}{N}\sum_{a_0,b_0}\langle0\vert\Phi(x)_{b_0a_0}[\Phi_{b_0a_0}(0)]^*\vert0\rangle\nonumber\\
&=&\frac{1}{N}\sum_{a_0,b_0}\sum_{\Psi}e^{ix\cdot p_{\Psi}}\langle0\vert \Phi(0)_{b_0a_0}\vert\Psi\rangle\langle\Psi\vert[\Phi(0)_{b_0a_0}]^*\vert0\rangle,\label{intermediatestates}
\eeq
where $\vert\Psi\rangle$ is any state with particles and antiparticles, and $p_\Psi$ is the sum of the momenta of the excitations of the state $\vert\Psi\rangle$.

After introducing the exact form factors in (\ref{intermediatestates}), the exact expression for the two-point function given in \cite{renormalizedfield} is
\beq
\mathcal{W}(x)=\frac{1}{4\pi}\sum_{l=0}^\infty\int_{-\infty}^{\infty} d\theta_1\dots\int_{-\infty}^{\infty} d\theta_{2l+1}\exp{\left(ix\cdot\sum_{j=1}^{2M-1} p_j \right)}\prod_{j=1}^{2l}\frac{1}{(\theta_j-\theta_{j+1})^2+\pi^2}.\label{correlationfinal}
\eeq
where the integration variables, $\theta_j$, correspond to the rapidities of the intermediate particles and antiparticles. These rapidities parametrize the energy and momentum of a particle by $E_j=m\cosh\theta_j,\,P_j=m\sinh\theta_j$.

For the purposes of this letter, we are interested only in the divergence at $x=0$ of (\ref{correlationfinal}). If we set $x=0$, the expression (\ref{correlationfinal}) diverges because one has to integrate over all the physical values of the rapidities. This divergence can be regularized by introducing a rapidity cutoff, $-\lambda<\theta_j<\lambda$. The rapidity cutoff, $\lambda$, is related to a standard Euclidean momentum cutoff, $\Lambda$, by 
\beq
 \lambda=\sinh^{-1}\left(\sqrt{\frac{\Lambda^2}{2m^2}-\frac{1}{2}}\right)=\ln\left(\sqrt{\frac{\Lambda^2}{2m^2}-\frac{1}{2}}+\sqrt{\frac{\Lambda^2}{2m^2}+\frac{1}{2}}\,\right)\approx \ln\left(\frac{\Lambda}{m}\right).\nonumber
\eeq
The regularized two-point function at $x=0$ is 
\beq
\mathcal{W}^\lambda(0)=\frac{1}{4\pi}\sum_{l=0}^\infty\int_{-\lambda}^{\lambda} d\theta_1\dots\int_{-\lambda}^{\lambda} d\theta_{2l+1}\prod_{j=1}^{2l}\frac{1}{(\theta_j-\theta_{j+1})^2+\pi^2}.\label{cutoffcorrelator}
\eeq

The short-distance behavior of this PCSM two point function was first studied in \cite{asymptoticfreedom}, and the regularized function at $x=0$ was shown in \cite{thooftthermodynamics}.  We will not show the full computation, but only quote the result we need, and refer the reader to the original paper. The result of \cite{asymptoticfreedom} and  \cite{thooftthermodynamics} is (after redefining integration variables $u_j=\theta_j/\lambda$)
\beq
\mathcal{W}^\lambda(0)=\frac{\lambda^2}{8\pi^2}\int\sum_{n=1}^\infty\left\vert\int_{-1}^{1} du \varphi_n(u)\right\vert^2\alpha_n^{-1}+\mathcal{O}(\lambda)=C_2\lambda^2+\mathcal{O}(\lambda).\label{logsquared}
\eeq
where $\alpha_n$ and  $\varphi_n(u)$ are the eigenvalues and eigenfunctions of the fractional Laplacian operator, $\Delta^{1/2}=\sqrt{-d^2/du^2}$. The upper bound $C_2<0.0219$
was found in \cite{asymptoticfreedom}\footnote{We have learned by private communication that Eytan Katzav and Peter Orland have calculated $C_2$ and found it is exactly $\frac{1}{16\pi}$. We  will only use the upper bound found in \cite{asymptoticfreedom}, as the calculation of the exact value has not yet been published.}. The correlation function diverges, being proportional to $(\ln\Lambda)^2$, which agrees with the perturbative result and asymptotic freedom \cite{polyakovbook}.

\section{Finite Volume Correlation Function}

The PCSM two-point function  at finite volume has been calculated in \cite{thooftthermodynamics}. This is done by compactifying the $x^1$ dimension into a circle of length $V$.  In 1+1 dimensions, this is equivalent to placing the system in a finite temperature $T=1/V$ (if we compactify $x^0$ instead of $x^1$).  Using the exact S-matrix, and the thermodynamic Bethe ansatz (TBA) \cite{thermodynamical} one can calculate the partition function. 

It was shown in \cite{thooftthermodynamics} that the the partition function of the PCSM coming from the TBA  at large $N$ is trivial and equivalent to a free bosonic gas. However, the correlation functions of operators are not those of the free theory. 

The two-point function of the renormalized field was evaluated at finite volume using the Leclair-Mussardo formula \cite{leclairmussardo}, which we will not discuss here. This is essentially a spectral sum, similar to (\ref{intermediatestates}), but where one sums not only over all the intermediate states of the two-point function, but one also includes an ensemble average, with a second sum over states weighted by the thermal distribution function.

There are several problems with the Leclair-Mussardo formula for two-point functions that make it not generally valid for all integrable theories \cite{saleur}.  The strongest objection is that generally, the expansion is not well defined, as there are poles for real values of the rapidities that need to be integrated over. However, it was argued in \cite{thooftthermodynamics} that these problems do not affect the PCSM at large $N$, and therefore the two-point function should be valid. For this model, the poles do not lie in the real line of rapidities, and the Leclair-Mussardo formula is well defined.

We are interested in the two-point function at $x=0$, but in a finite volume, $V$. This result is found in Eq. (6.6) of Ref. \cite{thooftthermodynamics}: 
\beq
&&\mathcal{W}^\lambda(0)^V=\frac{1}{4\pi}\sum_{l=0}^{\infty}\sum_{n^1=0}^{2l}\,\sum_{n^2=0}^{2l-n^1}\sum_{n^4=0}^{2l-n^1-n^2}\int_{-\lambda}^{\lambda}d\theta_{1}\dots \int_{-\lambda}^\lambda d\theta_{2l+1}\left[f_1(\theta_1)f_1(\theta_{2l+1})\right]^{\frac{1}{2}}\nonumber\\
&&\times\prod_{j=1}^{n^1}F_{1,\,1}(\theta_j,\theta_{j+1})\prod_{j=n^1+1}^{n^1+n^2}F_{1,\,-1}(\theta_j,\theta_{j+1})\prod_{j=n^1+n^2+1}^{n^1+n^2+n^4}F_{-1,\,-1}(\theta_j,\theta_{j+1})\prod_{j=n^1+n^2+n^4+1}^{2l}F_{-1,\,1}(\theta_j,\theta_{j+1})\nonumber\\
&&+\frac{1}{4\pi}\sum_{l=0}^{\infty}\sum_{n^1=0}^{2l}\,\sum_{n^2=0}^{2l-n^1}\sum_{n^3=0}^{2l-n^1-n^2}\int_{-\lambda}^{\lambda}d\theta_{1}\dots \int_{-\lambda}^\lambda d\theta_{2l+1}\left[f_{-1}(\theta_1)f_{-1}(\theta_{2l+1})\right]^{\frac{1}{2}}\nonumber\\
&&\times\prod_{j=1}^{n^3}F_{-1,\,1}(\theta_j,\theta_{j+1})\prod_{j=n^3+1}^{n^1+n^3}F_{1,\,1}(\theta_j,\theta_{j+1})\prod_{j=n^1+n^3+1}^{n^1+n^2+n^3}F_{1,\,-1}(\theta_j,\theta_{j+1})\prod_{j=n^1+n^2+n^3+1}^{2l}F_{-1,\,-1}(\theta_j,\theta_{j+1})
.\nonumber\\
&&\label{regularizedfinitevolumecorrelator}
\eeq	
where $F_{\sigma_i,\,\sigma_j}(\theta_i,\theta_j)=\frac{[f_{\sigma_i}(\theta_i)f_{\sigma_j}(\theta_j)]^{\frac{1}{2}}}{(\theta_i-\theta_j)^2+2^{\vert \sigma_i-\sigma_j\vert}\pi}$, with $\sigma_{i,j}=\pm 1$, and  $f_{\pm 1}(\theta_j)=1/(1+e^{\mp V m\cosh\theta_j})$. 

The expression (\ref{regularizedfinitevolumecorrelator}) is analyzed for both large and small values of $V$ in \cite{thooftthermodynamics}. At large $V$, one simply recovers the infinite volume expression (\ref{cutoffcorrelator}). The small-volume limit is studied by realizing that for small volumes (for $mV<<1$), the functions $f_{\pm 1}(\theta_j)$ become approximately
\beq
f_1(\theta)=\left\{\begin{array}{c}
\frac{1}{2},\,\,\,\,-\mathcal{L}<\theta<\mathcal{L},\\
\,\\
1,\,\,\,\,{\rm otherwise},\end{array}\right.\,\,\,,\,\,\,\,\,\,\,\,\,\,\,\,\,\,\,
f_{-1}(\theta)=\left\{\begin{array}{c}
\frac{1}{2},\,\,\,\,-\mathcal{L}<\theta<\mathcal{L},\\
\,\\
0,\,\,\,\,{\rm otherwise}.\end{array}\right.\,\,\,,\nonumber
\eeq
where $\mathcal{L}=\ln\frac{1}{mV}$.

Using techniques similar to those of \cite{asymptoticfreedom}, it was found \cite{thooftthermodynamics} that for extremely small volumes, with $\mathcal{L}\approx\lambda$,
\beq
W^\lambda(0)^V\approx C_2(\lambda^2-\mathcal{L}^2)+\frac{8}{\pi}\mathcal{L}+\mathcal{O}(\mathcal{L}^0)+\mathcal{O}(\lambda-\mathcal{L}).\label{softened}
\eeq
 At the smallest possible physical volume of $V=1/\Lambda$ the $(\ln\Lambda)^2$ term completely cancels out. For these very small volumes, the divergence of the two-point function is reduced from  order $\lambda^2$ to order $\lambda$.

\section{The Higgs phase and gluon mass}


We have presented the two-point correlation function of the renormalized field operator of the PCSM at large $N$ in a finite volume. For very large volumes,  the standard, infinite-volume two-point function from \cite{renormalizedfield},\cite{asymptoticfreedom}, which diverges as $(\ln\Lambda)^2$ at short distances, is recovered. For very small volumes, we saw that the order of this divergence reduces from  $(\ln\Lambda)^2$ to $\ln\Lambda$ as we reduce the volume size. 

We would like to discuss in more detail what we mean by very small volumes. Suppose the momentum cutoff originates from placing the theory on a lattice with spacing $a\sim 1/\Lambda$. Now suppose the length of the $x^1$ direction is $V=na$. For the lattice computations of Ref. \cite{gongyozwanziger}, for example, the values used were $n=256,\,512,\,1024$. As an arbitrary example, for the sake of illustrating our point, let's imagine a lattice spacing is chosen such that $ma=10^{-14}$ (this is the assumption that $m$ is small enough that all these volumes are still in the regime where $mV<1$). It will soon be clear why this particular value of $ma$ is chosen and is relevant. For this spacing, the rapidity cutoff is $\lambda=32.236$. We can calculate $\mathcal{L}$ for the different volumes $n=256,\,512,\,1024$, for which we find $\mathcal{L}_{256}=26.691,\,\mathcal{L}_{512}=25.998,\,\mathcal{L}_{1024}=25.305$. That is, for typical lattice volumes, like those of Ref. \cite{gongyozwanziger},  $\mathcal{L}$ and $\lambda$ are of the same order of magnitude. The ``softening" of the divergence from (\ref{softened}) is a large and significant effect for typical lattice computations at small volume.

The question remains, 
 is the PCSM mass gap, $m$, small enough that $mV<<1$ for the volumes studied in Ref. \cite{gongyozwanziger} ? The only length scale in the PCSM is $m$, and this parameter is usually fixed, and the bare coupling $g_0(\Lambda)$ varies as we change the cutoff, $\Lambda$. Their relationship is \cite{polyakovbook}
\beq
 \frac{m}{\Lambda}=ma=\frac{K}{g_0}\left(e^{-2\pi/g_0^2}+\dots\right),\label{massrenormalization}
 \eeq
 where $K$ is some non-universal constant, which depends on the regularization procedure. The parameter that was controlled by hand in Ref. \cite{gongyozwanziger} is $\gamma=1/g_0^2$. The values used were in the range $\gamma\in\{1,...,10\}$, with a crossover into the Higgs phase detected around $\gamma=5,6,7,8$. As is seen from Eq. (\ref{massrenormalization}), 
 the mass gap is very small for the values of $\gamma$ at which the crossover was seen (and thus the size of the PCSM particles is very large). For example, if we take $K=1$, we find $ma(\gamma=5)=5.07834\times10^{-14}$, and $ma(\gamma=8)=4.18334\times10^{-22}$. It is then clear that the computations of \cite{gongyozwanziger} are all done deep in the small-volume regime.

We stated that the only reason for the absence of a Higgs phase in the continuum theory is that  the gluon mass is proportional to the renormalized-field two-point function at $x=0$. The gluon mass becomes finite once a momentum cutoff is introduced by placing the theory on a lattice. Depending on the value of the coupling constants and the volume, this mass can become small enough to be observed in Monte Carlo simulations.

We can estimate the gluon mass using Eq. (\ref{gluonmass}) and  (\ref{softened}), with $C_2\approx 0.0219$. We compare our mass estimates with those of Table 2. of Ref. \cite{gongyozwanziger}. The gluon mass is a function of the variables, $g_0,\,e$ and $n=V/a$. We now switch to the notation of \cite{gongyozwanziger}, where the couplings used are $\gamma,$ and $\beta=1/e^2$. For small volumes, the gluon mass is approximately (with $K=1$)
\beq
\mathcal{M}(\gamma,\beta, n)=\left\{\frac{\gamma}{\beta}C_2 \left[\ln^2\left(\frac{1}{\sqrt{\gamma}}e^{2\pi\gamma}\right)-\ln^2\left(\frac{1}{n\sqrt{\gamma}}e^{2\pi\gamma}\right)\right]+\frac{\gamma\, 8}{\beta\,\pi}\ln\left(\frac{1}{n\sqrt{\gamma}}e^{2\pi\gamma}\right)\right\}^{\frac{1}{2}}.\label{massformula}
\eeq
Equation (\ref{massformula}) is not expected to agree quantitatively with the results of \cite{gongyozwanziger}, as it is valid only in the large-$N$ limit. However, we argue that it qualitatively explains some of the observed behavior.

First we examine the $\beta$ dependence. As can be observed in Table 2 of \cite{gongyozwanziger}, increasing the coupling, $\beta$ while keeping $\gamma$ and $n$ constant, generally decreases the gluon mass (this becomes clearer in their computations at higher values of $\gamma$, closer to the continuum limit). The clearest example shown in \cite{gongyozwanziger} is for $\gamma=8, \,n=256$, where they find $\frac{\mathcal{M}(\beta=120)}{\mathcal{M}(\beta=200)}=1.288$. Plugging these same parameters into (\ref{massformula}), we find $\frac{\mathcal{M}(\beta=120)}{\mathcal{M}(\beta=200)}=1.291$, which we believe confirms the $\beta$ dependence of the gluon mass.

The size of the lattice spacing relative to the PCSM mass gap is controlled solely by the $\gamma$ parameter (Eq. (\ref{massrenormalization})). The continuum limit is equivalent to the limit, $\gamma\to\infty$. It is clear from (\ref{massformula}) that the gluon mass diverges as expected in this limit. For high values of $\gamma$, keeping $\beta$ and $n$ fixed, we expect the gluon mass to increase as we increase $\gamma$. This is confirmed in \cite{gongyozwanziger}, where the clearest example is for their largest measured values $\gamma$, fixing $\beta=120$, $n=256$, finding $\frac{\mathcal{M}(\gamma=8)}{\mathcal{M}(\gamma=10)}=0.928$. From the formula (\ref{massformula}), for these values, we find $\frac{\mathcal{M}(\gamma=8)}{\mathcal{M}(\gamma=10)}=0.714401$. Here we expect that the large-$N$ limit plays a larger role, since the powers of the logarithms in (\ref{massformula}) will be different at smaller $N$.

Finally, we verify qualitatively the volume dependence of the gluon mass. As can be observed from Table 2 of \cite{gongyozwanziger}, how the gluon mass reacts to a change of volume depends on the value of $\gamma$. For small values of $\gamma$, increasing the volume, slightly decreases the mass. For example, fixing $\beta=120$, $\gamma=5$, they found $\mathcal{M}(n=256)= 0.240(2),$ and $\mathcal{M}(n=512)=0.237(3)$ (in lattice units). However there is a change in behavior somewhere between $\gamma=5$ and $\gamma=6$. For large values of $\gamma$, increasing the volume slightly increases the mass. For example, for $\beta=120,$ $\gamma=6$, they find $\mathcal{M}(n=256)=0.265(5)$, and $\mathcal{M}(n=512)= 0.267(2)$. Our formula, (\ref{massformula}) reproduces exactly this behavior. For a fixed value of $\beta$ and $\gamma$, the mass is modified very slightly by changing the volume from $n=256$ to $n=512$. For small $\gamma$, the mass decreases with increasing volume, and for large $\gamma$, the mass increases with volume. This change in behavior for Eq. (\ref{massformula}) happens around $\gamma\approx10.38$. To illustrate with some examples, for $\beta=120$, and $\gamma=5$, we have $\mathcal{M}(n=256)=1.71503$, $\mathcal{M}(n=512)=1.70266$. For $\gamma=12$, we have $\mathcal{M}(n=265)=4.38238,$ $\mathcal{M}(n=512)=4.38589.$

As was remarked in \cite{massiveyangmills}, for the value of $\gamma$ used in \cite{gongyozwanziger}, the PCSM mass gap, $m$, is very small, compared to the coupling $e$, and becomes smaller as $\gamma$ is increased. This means that it is very difficult to observe the confined phase, since string breaking occurs very easily. Since we have shown that there exist finite-mass gluons in the lattice theory at finite volume, it is reasonable to observe a Higgs-like potential between two sources at large separations, which explains the smooth crossover  of \cite{gongyozwanziger}.


We have shown that at infinite $N$, the gluon mass has qualitatively similar behavior to the one observed numerically in \cite{gongyozwanziger} for $N=2$. Now the very important question remains, what can we actually say analytically about the finite-$N$ case (particularly $N=2$)?

Reproducing a nonperturbative result like the correlation functions (\ref{logsquared}), (\ref{softened}) for general $N$ is a very difficult task that we are unable to achieve at this point. There are several difficulties to overcome. 

First of all, the form factors of the PCSM renormalized field at finite $N$ are not known, and it is significantly harder to calculate them. In fact, the simplicity of the S-matrix at large $N$ is what made the calculation of form factors possible.  A notable exeption is the $N=2$ case, where the PCSM is equivalent to an $O(4)$-symmetric nonlinear sigma model, using the fact that ${\rm SU}(2)\times{\rm SU}(2)\simeq O(4)$. Some of the few-particle form factors of the O(4) model have been calculated in \cite{ofour}. However, we need an explicit expression of all the form factors to study the ultraviolet regime of the correlation function, so the results of \cite{ofour} are not enough to complete the $N=2$ computation. 

Once all the form factors are known there is still the second problem of extracting the ultraviolet information from the correlation function by analyzing the spectral sum (\ref{intermediatestates}) as was done in \cite{asymptoticfreedom} for large $N$. This was a very nontrivial task that was possible only because of the simple structure of the large-$N$ form factors, where $\mathcal{O}(1/N)$ corrections were discarded at various points. It will very likely take a lot more time and effort to extract a simple result like (\ref{logsquared}) from the finite-$N$ form factors, even if the form factors were known.

Finally, if we want to study the finite-$N$ correlation function at finite volume/temperature, the simple Leclair-Mussardo formula is no longer valid, and it is not clear what is the way to proceed with this calculation. As was shown in \cite{thooftthermodynamics}, the Leclair-Mussardo formula is only expected to be valid at infinite $N$, because of two important properties: the expansion is well defined (there are no poles in the real line of rapidities), and the thermal distributions arising from the thermodynamic Bethe ansatz are trivial (like that of an ideal gas). There is no reason to believe these properties persist at finite $N$. It is likely that a more careful regularization scheme is needed, like the one proposed in \cite{takacs}. 

Having established the difficulty of extending our results to general $N$, one can consider the possibility of studying large, but not infinite $N$, by finding small corrections to our result in powers of $1/N$. In principle this could be possible, even though it might require a very difficult and careful calculation, since there are several steps of the computation where $\mathcal{O}\left(1/N\right)$ corrections have been ignored. To compute these small corrections, one needs to keep track of the terms that have been ignored at the levels of the S-matrix, computation of form factors, TBA, and correlation functions.

At the S-matrix level, it is very simple to find $1/N$ corrections, since and exact expression is known for all $N$ \cite{wiegmann}. However, the most important simplifying property of the infinite-$N$ limit is that, as was discussed in \cite{thooftthermodynamics}, the scattering becomes effectively diagonal. In this case this means that most particles don't interact with each other, unless there is a color-index contraction between them. Small $1/N$ corrections already present a difficulty when computing form factors, since it means that all particles interact with each other. There is no fundamental reason why this computation should not be possible, but there will be many more contributions to the form factors to take into account from new particle interactions that were not present at infinite $N$.

Our TBA and Leclair-Mussardo correlation-function computations were significantly simplified by relying on the fact that the scattering was diagonal. TBA computations are much more difficult in a non-diagonal theory, since one needs to take into account the thermodynamic contributions of pseudo particles, such as magnons and their bound states (also called "strings"). A full non-diagonal TBA computation for the PCSM was proposed in \cite{kazakovleurent}, but the resulting functional equations have not yet been solved for large values of $N$. The original proposal for computing correlations functions by Leclair and Mussardo was also only applicable to diagonal scattering theories. A proposal for generalizing the computation of these correlation functions to non-diagonal theories was made in Ref. \cite{buccheritakacs}.

There is in principle not a physical reason why it should not be possible to compute $1/N$ corrections. But as we have discussed, even a small correction would completely change the techniques we have to use, since the PCSM is a diagonal scattering theory only at precisely $N=\infty$.

Despite all these negative arguments, there are still some statements we can make about the Higgs phase at finite $N$. There are, however, two disadvantages: we are forced to rely only on perturbative calculations, and we can only study the infinite volume case. 

A simple perturbative analysis tells us how the general-$N$ two-point function diverges at $x=0$. This simple result can be found in the first reference of \cite{polyakovbook}, and is also discussed in \cite{asymptoticfreedom}. From leading-order perturbation theory one can find the time-ordered two-point function, $G(x,\Lambda)$, and the coupling $g_0(\Lambda)$ satisfy the renormalization group equations:
\beq
\frac{\partial\ln G(x,\Lambda)}{\partial \ln\Lambda}=\gamma(g_0)=\gamma_1 g_0^2+\cdots,\,\,\frac{\partial g_0^2(\Lambda)}{\partial \ln \Lambda}=\beta(g_0)=-\beta_1g_0^4+\cdots,\label{renormalizationgroup}
\eeq
with the coefficients $\gamma_1=(N^2-1)/(2\pi N^2)$ and $\beta_1=1/4\pi$. Integrating (\ref{renormalizationgroup}), one finds
\beq
G(0,\Lambda)=C \left[\ln\left(\frac{\Lambda}{m}\right)\right]^{\gamma_1/\beta_1}+\cdots,\label{twopointpolyakov}
\eeq
where $C$ is some undetermined constant. Equation (\ref{twopointpolyakov}) is completely consistent with the nonperturbative (\ref{logsquared}), since $\lim_{N\to\infty} \gamma_1/\beta_1=2$, which is the main result of Ref.\cite{asymptoticfreedom}. One advantage of the nonperturbative calculation is that one can find the constant $C=C_2$, while for finite $N$ this is unknown from this calculation.

This very simple perturbative result allows us to write an expression for any $N$ for the gluon mass at infinite volume, up to the constant $C$. Combining (\ref{twopointpolyakov}) with (\ref{massrenormalization}), we can write for $N=2$ (using the notation of Eq. (\ref{massformula}), for $n\to\infty$)
\beq
\mathcal{M}(\gamma,\beta)=\left\{\frac{\gamma}{\beta} C\left[\ln\left(\frac{1}{\sqrt{\gamma}}e^{2\pi\gamma}\right)\right]^{3/2}\right\}^{\frac{1}{2}}. \label{massformulafinite}
\eeq
Equation (\ref{massformulafinite}) is not good enough to compare with the results from \cite{gongyozwanziger} because as we argued, their computations are done in very small volumes. At small volumes we expect there might appear some volume-dependent contributions like those of (\ref{massformula}) that cannot be ignored.

Even though the result (\ref{massformulafinite}) is only valid for infinite volume, it can still show the basic conclusion of this paper. That is, introducing a lattice regularization makes the gluon mass finite, and even small, depending on the values of the couplings. For example, we show some values that span the range used in \cite{gongyozwanziger},  $\mathcal{M}(2,120)=0.843767\sqrt{C}$, \,\,\,$\mathcal{M}(10,120)=6.35361 \sqrt{C}$, \,\,\, $\mathcal{M}(2,200)=0.653579\sqrt{C}$, \,\,\,$\mathcal{M}(8,200)=3.71684\sqrt{C}$.

\acknowledgments

 I would like to thank Peter Orland and Ra\'ul Brice\~no for many helpful discussions. This work has been supported by the ERC, under grant number 279391 EDEQS.



\end{document}